\documentclass[twocolumn,showpacs,preprintnumbers,amsmath,amssymb]{revtex4}

\usepackage{graphicx}%
\begin{document}

\newcommand{\bc}{\begin{center}} \newcommand{\ec}{\end{center}}
\newcommand{\be}{\begin{equation}} \newcommand{\ee}{\end{equation}}
\newcommand{\beqn}{\begin{eqnarray}} \newcommand{\eeqn}{\end{eqnarray}}

\title{Strong Griffiths singularities in random systems and their relation to extreme value statistics}

\author{R\'obert Juh\'asz, Yu-Cheng Lin}
 \email{juhasz,lin@lusi.uni-sb.de} \affiliation{
 Theoretische Physik, Universit\"at des Saarlandes, D-66041
 Saarbr\"ucken, Germany  }%

\author{Ferenc Igl\'oi}  \email{igloi@szfki.hu}
\affiliation{ Research Institute for Solid
State Physics and Optics, H-1525 Budapest, P.O.Box 49, Hungary}
\affiliation{ Institute of Theoretical Physics, Szeged University,
H-6720 Szeged, Hungary}

\date{\today}

\begin{abstract}
We consider interacting many particle systems with quenched disorder having strong Griffiths singularities,
which are characterized by the dynamical exponent, $z$,
such as random quantum systems and exclusion processes. In several $d=1$ and $d=2$ dimensional problems we have
calculated the inverse time-scales, $\tau^{-1}$, in finite samples of linear size, $L$, either exactly or numerically. In all cases, having a discrete symmetry, the distribution function, $P(\tau^{-1},L)$,
is found to depend on the variable, $u=\tau^{-1}L^{z/d}$, and to be universal given by the limit distribution of extremes of independent and identically distributed random numbers. This finding is
explained in the framework of a strong disorder renormalization group approach when, after fast degrees of
freedom are decimated out the system is transformed into a set of non-interacting localized excitations.
The Fr\'echet distribution of $P(\tau^{-1},L)$ is expected to hold for all random systems having a
strong disorder fixed point, in which the Griffiths singularities are dominated by disorder fluctuations.

\end{abstract}

\pacs{Valid PACS appear here}

\maketitle
\section{Introduction}

In interacting many-particle systems quenched (i.e. time independent) disorder can induce
unusual physical properties which do not exist in non-random systems. For example in a
random ferromagnet the linear susceptibility, $\chi$, is divergent in an extended part
of the paramagnetic phase, which is called Griffiths phase\cite{griffiths}. This type of singular
behavior is caused by rare regions of the sample, in which due to
strong disorder fluctuations the local couplings are much stronger than their average and
they are even stronger than the disordering field, such as
the temperature in a classical system. Consequently these rare regions are locally in the
ferromagnetic phase and the relaxation process associated to them
involves a very large relaxation time, $\tau$. In the thermodynamic limit
the distribution of the $\tau$-s has no upper limit which leads to singularities in several
average physical quantities. One can say that these Griffiths singularities are
controlled by a line of semi-critical fixed points, having a diverging relaxation time but a
finite correlation length, $\xi$. In a classical disordered system in which the
phase transition is triggered by the variation of the temperature these singularities are
very weak\cite{harris} and often one needs astronomical times to be able to observe them. There are, however,
another type of random systems in which the Griffiths effects are much stronger and one
can measure them in conventional experiments.

Strong Griffiths effects can be observed, among others, in random quantum systems\cite{qsg} at zero temperature,
$T=0$, in which a quantum phase transition can take place by varying a control parameter, $\delta$.
The prototype of random quantum systems is the random transverse-field Ising model (RTIM), which
is defined in one dimension (1d) by the Hamiltonian,
\be
H= -\sum_{i=1}^{L-1} \lambda_i \sigma_i^x \sigma_{i+1}^x - \sum_{i=1}^{L} h_i \sigma_i^z\;,
\label{RTIM}
\ee
in terms of the Pauli matrices, $\sigma_i^{x,z}$, at site $i$. The exchange couplings, $\lambda_i$, and
the transverse fields, $h_i$, are both independent and identically distributed ({\it iid}) random variables
and the control parameter is defined\cite{fisher} as:
\be
\delta=\frac{[\ln h]_{\rm av} - [\ln \lambda]_{\rm av}}{{\rm var}[\ln \lambda] + {\rm var}[\ln h]}\;.
\label{delta}
\ee
Here and in the following we use $[ \dots ]_{\rm av}$ to denote averaging over quenched disorder
and ${\rm var}[x]$ stands for the variance of $x$. The system is in the ferromagnetic (paramagnetic) phase
for $\delta<0$ ($\delta>0$) and the quantum critical point is located at $\delta=0$.
It was first McCoy\cite{mccoy} who calculated exactly the average susceptibility of this model and found that
$[\chi]_{\rm av}$ is divergent in a finite regime of the paramagnetic phase. Later the excitation
energy, $\epsilon \sim 1/\tau$, is shown\cite{fisher} to vanish with the size of the system as:
\be
\epsilon \sim L^{-z}\;,
\label{z}
\ee
with the dynamical exponent, $z>0$, which is a continuous function of $\delta$.

Another class of problems with strong Griffiths effects are stochastic many-particle
systems\cite{schutzreview}) with quenched disorder\cite{krug}, such as the 1d partially asymmetric simple exclusion
process (PASEP)) with position or particle dependent hopping rates\cite{ASEP,pasep_sw} or the zero range process (ZRP) with disorder\cite{ZRP}. Also reaction-diffusion models\cite{hinrichsen} with quenched disorder might show strong Griffiths effects\cite{contact}, see, however\cite{vojta,odor}.
As an example we consider here the PASEP with particle-wise ({\it pw}) disorder\cite{ASEP}, in which the $i$-the particle,
$i=1,2,\dots,N$, can hop to empty neighboring sites with a rate $p_i$ to the right and $q_i$ to the
left in a 1d periodic lattice with $L > N$ sites. Here the hop rates are {\it iid} random numbers and the
control parameter is defined as:
\be
\delta_p=\frac{[\ln p]_{\rm av} - [\ln q]_{\rm av}}{{\rm var}[\ln p] + {\rm var}[\ln q]}\;,
\label{delta_p}
\ee
so that the particles move to the right (to the left) for $\delta_p>0$ ($\delta_p<0$).
Here we restrict ourselves to the domain, $\delta_p \ge 0$.
Strong Griffiths effects in the PASEP are caused by such rare particle clusters in which the
local hopping rates of the particles prefer jumps against the global bias. As a result
in the anomalous diffusion regime, which can be called as the Griffiths phase of the PASEP,
the average displacement of particles grows in time as 
\be
\langle x (t)\rangle \sim t^{1/(1+z_p)},
\ee
when the system undergoes a coarsening process.
Here, the dynamical exponent of the PASEP, $z_p>0$, is a continuous function of $\delta_p$.
Note that outside the Griffiths phase,
$\delta > \delta_G$, the stationary velocity of the particles is $v=O(1)$, whereas in the
Griffiths phase it vanishes as\cite{ASEP}:
\be
v \sim L^{-z_p}\;.
\label{v}
\ee

In general Griffiths singularities are characterized by the distribution of the relaxation
times of slow processes or equivalently by the distribution of their inverse, $1/\tau$,
which is just the excitation energy, $\epsilon$, of the RTIM and the stationary velocity, $v$, of the PASEP.
In a finite system $1/\tau$ is just the smallest event associated to one of the rare regions
and the occurrence of these rare regions as well as the distribution of $1/\tau$ depends on
the form of the randomness in the dynamical model. In the mathematical literature such questions
are studied in the frame of extreme value statistics\cite{galambos} (EVS) and rigorous
results are known for {\it iid} random variables, which are distributed according to a
parent distribution, $\pi(y)$ ($y=1/\tau$). In this case the limit distributions are in a few different universal scaling forms depending on the asymptotic behavior of the parent distribution for large values of $y$. If the decay of $\pi(y)$ is faster than any power then it is the Gumbel distribution,
for power-law tails it is the
Fr\'echet distribution and for power-laws with an edge the Weibull distribution is the result.
If the random variables are not {\it iid} there are no general mathematical
results, however a few special cases have been studied and solved\cite{galambos}. Among others we can mention
the statistics of extreme intensities of Gaussian interfaces\cite{racz} which generally does not coincide with
the known limit distributions of {\it iid} variables. Similarly, for hierarchically correlated variables the
limit distribution deviates from the Gumbel form\cite{dean}.

As we described above strong Griffiths affects are present in interacting particle systems having strong
temporal and spatial correlations, therefore at first thought it seems unlikely that a simple connection with the theory of EVS exists. A closer look, however, shows that in a class of systems the rare events, which are
the source of Griffiths singularities, are well localized and separated in such a way that they could be
considered independent. For these systems then the distribution function, $P_L(1/\tau)$, could correspond
to the appropriate limit distribution of EVS. 

In this paper we are going to study this issue in more details. For some random 1d models having a
small correlation length we perform exact calculations. On the other hand at a general point of the
Griffiths phase a renormalization group (RG) approach\cite{im} is applied, which is expected to give
asymptotically exact dynamical singularities in the Griffiths phase. The RG equations can be analytically
solved for some 1d random models, such as for the RFIM\cite{fisher} and for the PASEP\cite{ASEP},
in other cases and in
higher dimensions one resort to numerical calculations. For some 1d random models we also calculate
numerically the different distribution functions.

The structure of the paper is the following. In Sec.\ref{sec:quantum} we consider Griffiths singularities
in random quantum systems. In more details we study the 1d RTIM, but also the random quantum Potts chain, as
well as the RTIM for ladders and in 2d and the random Heisenberg model in 1d and 2d are investigated. In Sec.\ref{sec:stochastic} we study Griffiths singularities in random stochastic systems, such as in the PASEP with particle-wise and
site-wise disorder. Our paper is closed by a discussion
in Sec.\ref{sec:disc}.

\section{Griffiths singularities in random quantum systems}
\label{sec:quantum}
\subsection{The RTIM in 1d}
\label{sec:RTIM}
\subsubsection{Exact results for extreme disorder}
\label{sec:RTIM_extr}
In order to gain some insight into the origin of Griffiths singularities we solve exactly
the RTIM in Eq.(\ref{RTIM}) using the bimodal distribution: $\lambda_i=\lambda$, with probability $c$
and $\lambda_i=\lambda^{-1}$ with probability $1-c$, and $h_i=1$. Furthermore we take the limit
$c \ll 1$, when most of the couplings are weak and $\lambda \gg 1$, so that
we are deeply in the paramagnetic phase. There
are, however, rare regions in which all the couplings, say $n$, are strong and appear
with an exponentially small density, $\rho(n)=c^n$. Such a cluster is typically
embedded into a see of very weak couplings, thus the corresponding excitation energy is obtained
by solving the gap in a cluster with free boundary conditions, leading to: $\epsilon(n) \approx \lambda^{-n}$.
Now we obtain that the distribution of the excitation energies in an infinite system
has a power-law tail:
\be
P(\epsilon) \approx \frac{1}{\ln \lambda} \epsilon^{\omega},\quad \epsilon \to 0\;,
\label{eps_extr}
\ee
and the gap exponent, $\omega$, is given by:
\be
\omega=\frac{\ln (1/c)}{\ln \lambda}-1\;.
\label{omega_extr}
\ee
In a finite system of size, $L$, the typical size of the largest cluster, $n_1$, is given
by the condition, $L \sum_{n \ge n_1}\rho(n)=1$, so that the typical value of the smallest gap,
$\epsilon_1$, is given by: $\epsilon_1=L^{-z}$, with the dynamical exponent defined in Eq.(\ref{z}):
\be
z=\frac{\ln \lambda}{\ln (1/c)}\;.
\label{z_extr}
\ee
Note that we obtain the relation, $\omega=1/z-1$, so that the distribution function for
finite systems satisfies the scaling form\cite{yr96}:
\be
P_L(\epsilon_1)=L^z \tilde{P}_1(\epsilon_1 L^{z})\;,
\label{P_L}
\ee
and the subscript of the scaling function, $\tilde{P}_1(u)$, refers to the first gap. In the
following we determine the scaling function. First, we note that the rare regions in the system are
localized thus they could be placed at $\sim L$ different positions of the chain. Furthermore
the different rare regions are independent and their lengths, which are proportional to the log-gap, is distributed by the same exponential distribution. From these follows that $\epsilon_1$ is the smallest gap
out of $\sim L$ independent possible gaps, associated to the different rare regions, and the
gaps are distributed by the same parent distribution given in a power-law form in Eq.(\ref{eps_extr}).
Consequently in the large-$L$ limit the distribution function, $\tilde{P}_1(u)$, is given by the
standard Fr\'echet distribution\cite{galambos}:
\be
\tilde{P}_1(u)=\frac{1}{z} u^{1/z-1} \exp(-u^{1/z})\;,
\label{frechet}
\ee
and $u=u_0 \epsilon_1 L^{z}$, where the non-universal constant, $u_0$, depends on the amplitude
of the tail in Eq.(\ref{eps_extr}).

\subsubsection{Analytical results of the strong disorder RG}
\label{sec:RTIM_RG}
The strong disorder RG method\cite{im} has been introduced by Ma and Dasgupta\cite{mdh} to study
random antiferromagnetic Heisenberg chains. For the 1d RTIM Fisher\cite{fisher} has applied the method
and solved analytically the RG equations in the vicinity of the random quantum critical point.
During renormalization the strongest local term in the Hamiltonian, coupling or transverse field,
is decimated out and between remaining sites new interactions are generated through perturbation
calculation. If a strong bond, say $\lambda_2$, connecting sites $1$ and $2$ is decimated out the effective
transverse field, $\tilde{h}$, acting on the two-site cluster is given by\cite{fisher}:
\be
\tilde{h}=\frac{h_1 h_2}{\kappa \lambda_2}\;,
\label{htilde}
\ee
where $\kappa =1$.
On the contrary if a strong transverse field, say $h_2$ and thus the site $2$ is decimated out the effective
coupling, $\tilde{\lambda}$, connecting the remaining sites, $1$ and $3$ is given by\cite{fisher}:
\be
\tilde{\lambda}=\frac{\lambda_1 \lambda_2}{\kappa h_2}\;,
\label{Jtilde}
\ee
with $\kappa =1$.
As renormalization goes on the energy scale, set by the strength of the strongest local term in the Hamiltonian, gradually decreases and the distributions of the renormalized couplings and transverse
fields approach their fixed-point form. At the critical point couplings and transverse fields
are decimated symmetrically and their fixed-points distribution functions are identical and have
the so called infinite disorder property\cite{fisher}: the ratio of any two terms tends either to zero or to
infinity. Consequently the decimation rules become exact and the critical singularities, both
static and dynamical ones are asymptotically exact. In the disordered Griffiths phase, which is
a part of the paramagnetic phase after a starting period of the RG when spin clusters of typical size,
$\xi$, are formed almost exclusively transverse fields are decimated out. At the fixed point the
energy scale still goes to zero and the decimation equations are asymptotically exact leading
to exact dynamical singularities, however the
spatial correlations are short-ranged and correct only up to a range of $\xi$. 
The analytical solution of the RG equations by Fisher\cite{fisher} is exact up to $O(\delta)$ which is
then extended into the complete
Griffiths phase\cite{ijl,i02}. The dynamical exponent, $z$, is found to be the positive root of the equation\cite{i02}:
\be
\left[ \left( \frac{\lambda}{h} \right)^{1/z}\right]_{\rm av}=1\;,
\label{z1}
\ee
which has also been derived by a mapping with the random random walk\cite{ir98}. For the binary disorder used in the previous section this leads to: $\lambda^{1/z}c
+\lambda^{-1/z}(1-c)=1$, having the solution in Eq.(\ref{z_extr}) in the limit $c \ll 1$. Note
that this result is valid for any value of $\lambda >1$, which is connected to the fact that the
gap of a large $n$-bond cluster in a see of weaker bonds scales as: $\epsilon(n) \sim \lambda^{-n}$.
The dynamical exponent in Eq.(\ref{z1}) is a continuous function of $\delta$, and in the
vicinity of the critical point it is divergent as\cite{fisher} $z \approx 1/2\delta$. It can be shown
by the RG calculation that the low temperature behavior of
the average susceptibility and that of the specific heat, $c_V$, are given by\cite{fisher,i02}:
\be
\chi(T) \sim T^{-1+1/z}, \quad c_V(T) \sim T^{1/z}\;,
\label{exp_T}
\ee
thus the singularity involves the dynamical exponent. Note that at zero temperature $\chi(0)$
is divergent for $z > 1$.

The distribution of the smallest energy gap in a finite chain of length, $L \gg \xi$, and in the
vicinity of the critical point: $0 < \delta \ll 1$ is calculated by Fisher and Young\cite{fisheryoung}. 
For the distribution of the log-gap, $G=-\ln \epsilon_1$, they obtained:
\be
d {\rm Prob}[G|L] \approx 2 \delta L n_G \exp(-L n_G) d G\;,
\label{prob_G}
\ee
where
\be
n_G \approx 4 \delta^2 \exp(-2\delta G)\;
\label{n_G}
\ee
is the density of remaining clusters. Here making use of the fact that in the vicinity of the critical point
the correlation length is given by\cite{fisher} $\xi \sim \delta^{-2}$ and $z \approx 1/2\delta$ we obtain:
\be
n_G \sim \frac{\epsilon_1^{1/z}}{\xi}\;.
\label{n_G1}
\ee
Inserting Eq.(\ref{n_G1}) into Eq.(\ref{prob_G}) we obtain that the scaling variable is $u=u_0 \epsilon_1 (L/\xi)^{z}$, in terms of which the distribution is in the Fr\'echet form in Eq.(\ref{frechet}). 

\subsubsection{Phenomenological considerations}
\label{sec:pheno}
Results of the previous two subsections indicate that the distribution of the smallest gap of the
RTIM in 1d is the Fr\'echet distribution, at least in the limiting cases i) extreme binary disorder
with $\xi \ll 1$ but $z$ is arbitrary (Sec.\ref{sec:RTIM_extr}) ii) arbitrary form of disorder,
but $\delta \ll 1$ (Sec.\ref{sec:RTIM_RG}). Here we argue that the Fr\'echet form of the
distribution should be generally valid in the disordered Griffiths phase, at least if the low-energy
excitations in the system are localized. Here we note that performing the RG up to an energy scale, 
$\Omega_{0}$, which is much smaller than $\Omega_{\xi}\sim \xi^{-z}$, we obtain an equivalent random spin chain with very weakly interacting effective spins. 
Typically, the log-bonds are in the order of $-\ln \tilde{\lambda} \sim \Omega_0^{-1/z}$, so they are small compared to the value of effective fields $\tilde{h} \Omega_0$\cite{i02}. Consequently, in the following RG steps almost exclusively transverse-fields 
are decimated out through the transformation in Eq.(\ref{Jtilde}), which  does not influence
the transverse fields of the other active spins.
At the energy scale, $\Omega_{0}$, the effective transverse fields, $\tilde{h}$, have a power-law distribution
$P(\tilde{h}) \sim \tilde{h}^{-1+1/z}$, and the smallest one gives thus the smallest gap of the chain.
Consequently the renormalized chain satisfies the conditions needed for the validity of the Fr\'echet limit distribution. 
Generally, for the $n$-th excitation energy the limit distribution is given by\cite{galambos}:
\be
\tilde{P}_n(u_n)=\frac{1}{z} u_n^{n/z-1} \exp(-u_n^{1/z})\;,
\label{frechet2}
\ee
in terms of $u_n=u_0 L^z \epsilon_n$. We note that the gap exponent is now, $\omega_n=n/z-1$,
which has been obtained before\cite{ijr} through scaling considerations.

This type of reasoning for the smallest gap applies to another random systems, too, provided the decimation rules
are analogous to those in Eqs.(\ref{htilde}) and (\ref{Jtilde}). In d-dimensions
the only difference is that $z$ is replaced by $z/d$,
provided the strong disorder RG approach leads to localized, non-interacting effective degrees of freedom.

Next, the distribution of excitation energies is analyzed in the ordered phase of the RTIM.
In $1d$ due to duality the disordered and ordered phases are related to each other and in the
RG procedure the role of the couplings and the transverse-fields are exchanged. Consequently,
at the energy-scale, $\Omega_0 \ll \Omega_{\xi}$, the typical log-fields are
$-\ln \tilde{h} \sim \Omega_0^{-1/z}$, whereas $\tilde{\lambda} \sim \Omega$. Thus the
renormalized chain is a classical random bond Ising chain with very small effective transverse fields having
an exponentially vanishing first gap. The dynamical properties of the chain are governed by kink-like
excitations\cite{huse}. In an open chain the lowest excitation is one kink, which corresponds through duality
to a spin flip in the disordered phase. Consequently the distribution of the second gap in an open chain
follows the Fr\'echet distribution. On the contrary for a closed chain the lowest excitation involves
two kinks and the corresponding excitation energy is the sum of the first two smallest effective couplings
of the renormalized chain. Consequently the distribution of the second gap in a closed chain is related
to the solution of the above extreme value problem. In higher dimensions the singularities
in the ordered Griffiths phase have a more complicated structure\cite{motrunich}. Here the
renormalized system is a
random bond Ising model with a complicated topology, and the (second) excitation, which is relevant
in the dynamical properties corresponds to the creation of an oppositely magnetized domain. The
excitation energy is just the sum of the effective couplings at the boundary of the domain.

In the following we check the above conjectures by numerical calculations.

\subsubsection{Distribution of gaps from numerical diagonalization}

\begin{figure}[!h]
\begin{center}
\includegraphics[width=3.2in,angle=0]{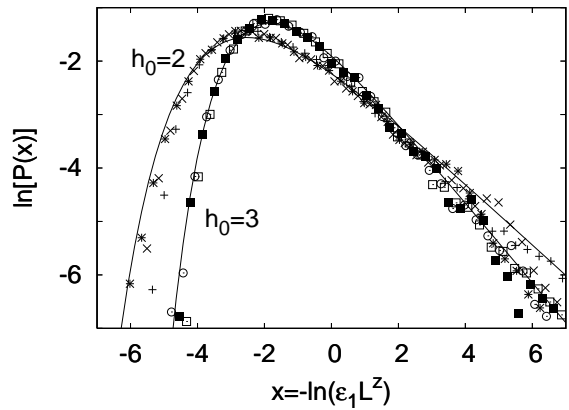}
\end{center}
\caption{Scaling plot of the distribution of the first gap of the 1d RTIM for different
sizes, $L=64$ ($+,\Box$), $128$ ($\times$,O) and $256$ ($\ast ,\blacksquare $) at two values of the uniform disorder. The full lines are the Fr\'echet distributions having
the exact values of the dynamical exponent, $z$.}
\label{fig:gap_RTIM}
\end{figure}

The RTIM in 1d can be transformed into a problem of free fermions and the calculation of the
gaps necessities the diagonalization of a $2L \times 2L$ tridiagonal matrix the entries of which
are the couplings and the transverse fields\cite{bigpaper}. In the numerical calculation we used a continuous,
uniform distribution: $\pi_{\lambda}(\lambda)=1$ for $0<\lambda\le 1$ and $0$ otherwise; as well as
$\pi_h(h)=1/h_0$, for $0 < h \le h_0$ and $0$ otherwise. The critical point is located at $h_0=1$,
the disordered Griffiths-phase is in the region $1 < h_0 < \infty$, and the dynamical exponent is given
by the solution of the equation: $z \ln(1-z^{-2})=- \ln h_0$. We have calculated the first gaps
for finite systems with $L=64,~128$ and $256$ and for two values of $h_0=2$ and $3$. The probability distribution of the first gaps are shown in Fig.\ref{fig:gap_RTIM} which all fit well to the Fr\'echet distribution.


\begin{figure}[!h]
\begin{center}
\includegraphics[width=3.2in,angle=0]{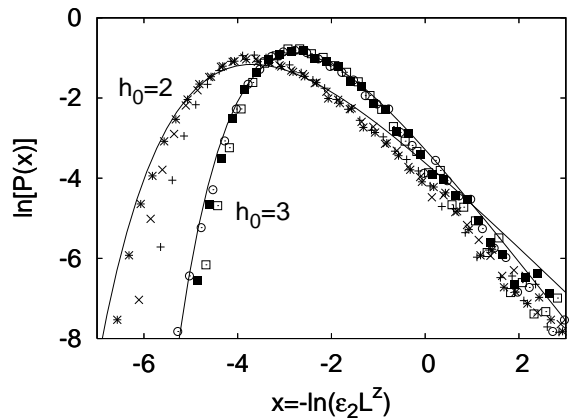}
\end{center}
\caption{The same as in Fig.\ref{fig:gap_RTIM} for the second gap.}
\label{fig:gap2_RTIM}
\end{figure}

The same conclusion is obtained with the distribution of the second gap, $\epsilon_2$, which is
presented in Fig.\ref{fig:gap2_RTIM}.


\subsection{Numerical RG study of chains, ladders and higher dimensional systems}

For more general models, in particular for a topology which is more complex than a linear chain
to calculate the gaps one resorts to numerical implementation of the strong disorder RG method.
This type of method is first used in Ref.\cite{motrunich} and implemented for a finite system
in Ref.\cite{lin00}. In the finite lattice version we use here
the decimation procedure is performed up to the last spin and the gap of the system
is identified by the effective transverse field
acting on that spin. By this procedure the first few RG steps are approximative, which will influence the
value of the non-universal constant, $u_0$, but the later transformation steps close to the fixed point are
presumably asymptotically exact. In the numerical calculations we used the uniform distribution and
considered at least $10000$ independent realizations of disorder.

\subsubsection{Random quantum Potts chain}

A simple generalization of the RTIM for $q$-state spin variables: $|s_i \rangle=|1 \rangle=|2 \rangle=
\dots |q \rangle$ is the random quantum Potts model defined in 1d by the Hamiltonian:
\be
H_P=-\sum_{i=1}^{L-1} \lambda_i \delta(s_i, s_{i+1}) - \sum_{i=1}^{L} \frac{h_i}{q}
\sum_{k=1}^{q-1} M_i^k\;,
\label{Potts}
\ee
where: $M_i|s_i \rangle=|s_i+1, {\rm mod}~q \rangle$. The control parameter of the model is in
the same form as for the RTIM in Eq.(\ref{delta}). The strong disorder RG approach is used for
this model in Ref.\cite{senthil} for the quantum critical point whereas in Ref.\cite{ijl} for the Griffiths phase. 
The transformation rules for the bonds and external fields are of the form given in Eqs.(\ref{Jtilde}) and (\ref{htilde}) with $\kappa=2/q$, thus the phenomenological argumentation directly apply here. The
only difference comparing with the RTIM is the degeneracy of the excited states. 
The distribution of the first gap is thus still of Fr\'echet type, and gaps between subsequent multiplets behave as the higher excitations of the RTIM. For the distribution of
the gaps in the ordered Griffiths phase one can obtain similar conclusions as described in Sec.\ref{sec:pheno}.

\begin{figure}[!h!]
\begin{center}
\includegraphics[width=3.2in,angle=0]{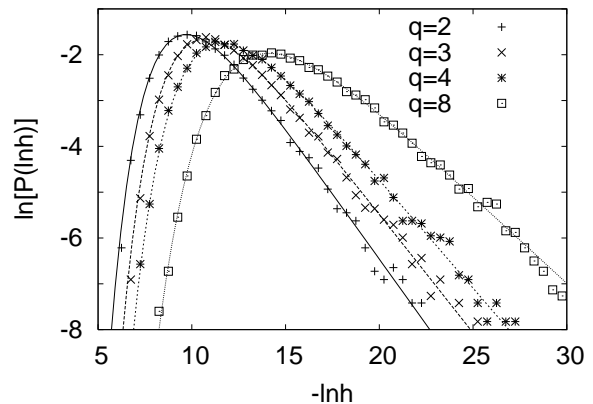}
\end{center}
\caption{Distribution of the log-gaps, identified as the transverse field of the last remaining
spin in the RG procedure, for the the random quantum Potts chain with different values of $q$
having a uniform distribution with $h_0=3.$.
The Fr\'echet distributions are indicated by full lines for which the dynamical exponent is obtained
by the solution of the analytical RG equations.}
\label{fig:potts}
\end{figure}

In the numerical application of the RG technique
we have calculated the distribution of the gaps at a large finite system, $L=2048$, for the uniform
distribution with $h_0=3$ but for different values\cite{XXX} of $q=2,3,4$ and $8$. The dynamical exponent
is $q$ dependent and calculated by a numerical integration of the analytical RG equations\cite{juhasz}. In
Fig.\ref{fig:potts} the numerically obtained gap distributions are compared with the Fr\'echet distribution
and excellent agreement is found.


\subsubsection{RTIM: ladder and 2d system}

Here we consider the RTIM with a more complex topology, first a ladder composed of two chains and 
afterwards a large plaquette of a square lattice. 

\begin{figure}[!h]
\begin{center}
\includegraphics[width=2.7in,angle=0]{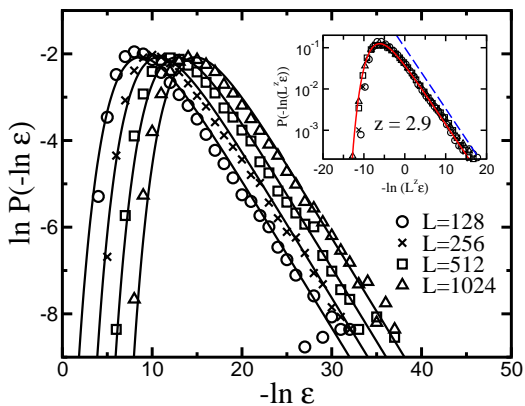}
\end{center}
\caption{(Color online) Distribution of the log-gaps, identified as the transverse field of the last remaining
spin in the RG procedure, for the ladder model with $h_0=2.5$ and for different lengths. In the
inset a scaling collapse according to Eq.(\ref{P_L}) with $z=2.9$ is presented. The Fr\'echet distributions
are indicated by full lines. The asymptotic
slope of the curves is indicated by a dashed line and is given by $\omega+1=0.34$, so that the relation in Eq.(\ref{omega_z}) is well satisfied.}
\label{fig:ladder}
\end{figure}

In Fig.\ref{fig:ladder} the distribution functions of the log-gaps of the ladder model with $h_0=2.5$ are presented
in a log-log scale for different lengths up to $L=1024$. The curves for different $L$-s are shifted to
each other and
a good scaling collapse can be obtained by using the scaling combination in Eq.(\ref{P_L}) 
with a dynamical exponent, $z=2.9$, as illustrated in the inset of Fig.\ref{fig:ladder}. The
(absolute value of the) asymptotic slope of the curves for small gaps is given by $\omega+1$, which
for localized excitations should be related to the dynamical exponent as:
\be
\omega+1=\frac{d}{z}\;,
\label{omega_z}
\ee
for a $d$-dimensional system. Here we obtained $\omega+1=0.34$ so that the relation in Eq.(\ref{omega_z})
with $d=1$ is very well satisfied. Finally, the scaled curve in the inset of Fig.\ref{fig:ladder}
is very well described by the Fr\'echet distribution.


\begin{figure}[!h]
\begin{center}
\includegraphics[width=2.7in,angle=0]{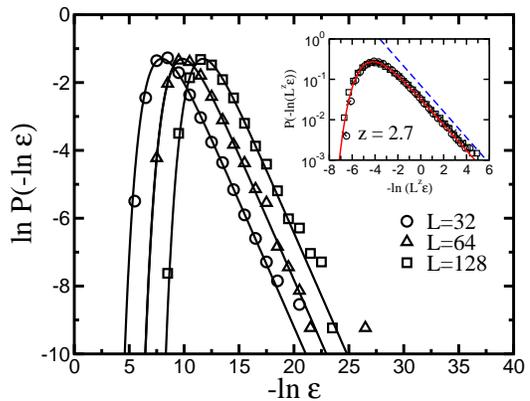}
\end{center}
\caption{(Color online) The same as in Fig.\ref{fig:ladder} for the square lattice RTIM with $h_0=9.$ and for different sizes.
In the inset the optimal scaling collapse is obtained with $z=2.7$. The asymptotic
slope of the curves is given by $\omega+1=0.74$, so that the relation in Eq.(\ref{omega_z}) is well satisfied.}
\label{fig:2dRTIM}
\end{figure}

Results of the same analysis of the gaps of the square-lattice RTIM for $h_0=9.$ are presented in
Fig. \ref{fig:2dRTIM} for $L=32$, $64$ and $128$. Here the scaling collapse in the inset is
obtained with $z=2.7$, whereas the gap exponent is given by $\omega+1=0.74$, so that the localization
condition in Eq.(\ref{omega_z}) is satisfied. Also the Fr\'echet distribution fits very well the
scaled curves.


\subsubsection{Random Heisenberg models in 1d and 2d}

The random quantum models studied so far have a discrete symmetry. In this section we are going
to consider random Heisenberg models which have continuous symmetry and the renormalization procedure
and the corresponding fixed points are somewhat different than for discrete symmetry models. 
To be specific the Heisenberg models we study here are defined by the Hamiltonian:
\be
H_H=\sum_{i,j} J_{i,j}t_i t_j \vec{S}_i \cdot \vec{S}_{j}\;,
\label{eq:SG}
\ee
in terms of spin-$1/2$ variables. $\vec{S}_i$ at site, $i$, and the dilution variables: $t_i=0$ with
probability, $p$, and $t_i=1$, otherwise. Here we consider two different types of models.
i) For the non-diluted models with $p=0$,
the random couplings are both ferromagnetic and antiferromagnetic and their average is $[J]_{\rm av}=0$. ii) For the diluted models we have $0<p<1$ and the random couplings are only antiferromagnetic, $J_{i,j}>0$. 

We start with non-diluted models in $1d$ and $2d$ and calculate the smallest gap in the system
by the numerical application of the RG method. As described in detail in Ref.\cite{lmri03} for these models
the RG scales into a so called large spin fixed point\cite{westerberg}, having an effective moment, $S_{eff} \sim L^{d \zeta}$,
with $\zeta \approx 1/2$. Due to the formation of large spins the low-temperature singularities
of these models are also different than that of the RTIM, for example the average susceptibility
has a Curie-like behavior.

\begin{figure}[!h]
\begin{center}
\includegraphics[width=2.7in,angle=0]{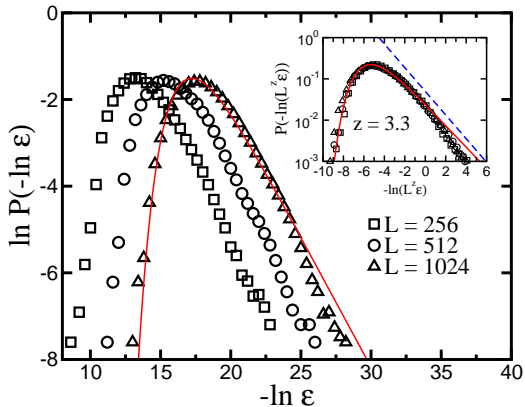}
\end{center}
\caption{(Color online) Distribution of the log gaps, obtained at the last step of the RG procedure
for the $1d$ non-diluted Heisenberg model
with uniform disorder, $-0.5 < J_i < 0.5$, and for different sizes.
In the inset the optimal scaling collapse is obtained with $z=3.3$. For small gaps
there is a considerable deviation from the Fr\'echet distribution, shown by a full line.
The asymptotic slope of the curves is given by $\omega+1=0.67$, so that the relation in Eq.(\ref{omega_z}) is not satisfied.}
\label{fig:1dSG}
\end{figure}

The distribution of the log-gaps for the $1d$ non-diluted model is shown in Fig. \ref{fig:1dSG} for different
sizes, and in the inset the scaled curves are presented with $z=3.3$. There are considerable
deviations from the Fr\'echet distribution, in particular for small gaps. The relation in Eq.(\ref{omega_z})
is not satisfied, so that the excitations are non localized.


\begin{figure}[!h]
\begin{center}
\includegraphics[width=2.7in,angle=0]{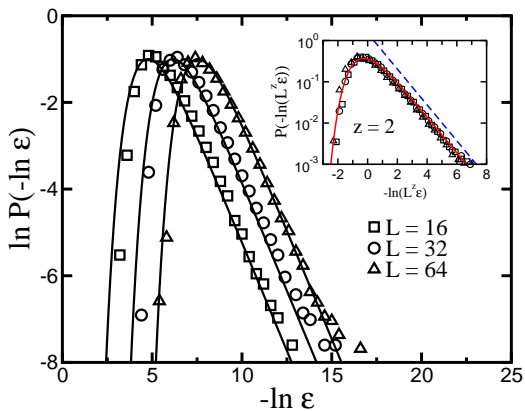}
\end{center}
\caption{(Color online) The same as in Fig.\ref{fig:1dSG} for the square lattice Heisenberg spin glass
with Gaussian disorder of variance $1$ for different sizes.
In the inset the optimal scaling collapse is obtained with $z=2$. The Fr\'echet distribution
is indicated by a full line and has a satisfactory fit to the data. The asymptotic
slope of the curves is given by $\omega+1=1$, so that the relation in Eq.(\ref{omega_z}) is satisfied.}
\label{fig:2dSG}
\end{figure}

The distribution of the log-gaps for the 2d system is shown in Fig. \ref{fig:2dSG} for three sizes,
$L=16,~32$ and $64$ and in the inset the scaled curves are presented with $z=2$. The Fr\'echet distribution
seems to give a correct description and also the the relation in Eq.(\ref{omega_z}) is satisfied. We
note that in 2d there is frustration in this model, thus it is called as a spin glass.


\begin{figure}[!h]
\begin{center}
\includegraphics[width=2.7in,angle=0]{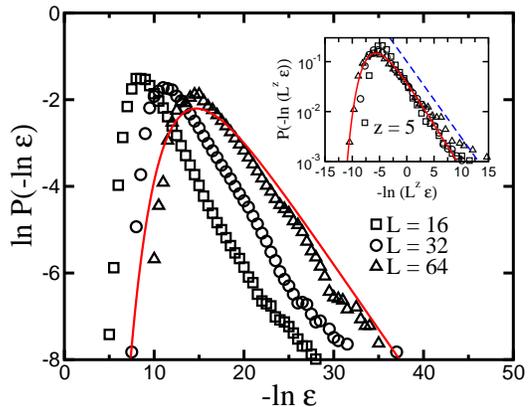}
\end{center}
\caption{(Color online) The same as in Fig.\ref{fig:1dSG} for the square lattice Heisenberg antiferromagnet
with dilution $p=0.125$.
In the inset the scaling collapse indicates $z=5$. The Fr\'echet distribution,
shown by a full line, differs considerably from the numerical data. The asymptotic
slope of the curves is given by $\omega+1=0.35$, so that the relation in Eq.(\ref{omega_z}) is satisfied
only with some error.}
\label{fig:2ddil1}
\end{figure}

\begin{figure}[!h]
\begin{center}
\includegraphics[width=2.7in,angle=0]{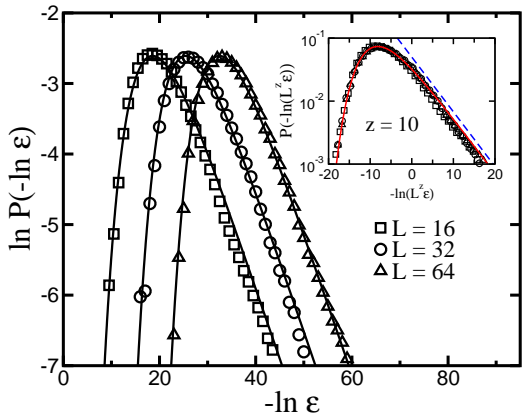}
\end{center}
\caption{(Color online) The same as in Fig.\ref{fig:2ddil1} for dilution $p=0.45$, above the percolation threshold.
In the inset the optimal scaling collapse is obtained with $z=10$. The Fr\'echet distributions,
indicated by full lines, fit very well the numerical data. The asymptotic
slope of the curves is given by $\omega+1=0.2$, so that the relation in Eq.(\ref{omega_z}) is satisfied.}
\label{fig:2ddil2}
\end{figure}

Next, we consider diluted models on the square lattice for which the couplings are antiferromagnetic
and distributed uniformly, $0 < J_{i,j} < 1$. For the dilution, $p$, we consider two values. In Fig.\ref{fig:2ddil1} we present the distribution
of the gaps at $p=0.125$, which is below the percolation threshold, whereas in Fig.\ref{fig:2ddil2}
we consider the dilution at $p=0.45$, above the percolation threshold, when the system is broken
into non-interacting finite clusters.

In the first case during renormalization the system scales into a large spin fixed point. As seen in Fig.\ref{fig:2ddil1} the distribution of the calculated gaps differs considerably from the Fr\'echet distribution and also the the relation in Eq.(\ref{omega_z}) is not correctly satisfied.


For $p=0.45$ the system is separated into independent parts and the smallest gap of the system is
just the smallest gap of these clusters. Since the energy gaps of the clusters are expected to
be distributed in identical power-law form, the applicability of the EVS is probable. Indeed, in
Fig.\ref{fig:2ddil2} the numerically calculated gap distributions can be well described by the
Fr\'echet form and also Eq.(\ref{omega_z}) is satisfied.


\section{Griffiths singularities in random stochastic systems}
\label{sec:stochastic}
\subsection{PASEP with extreme disorder}
\label{sec:PASEP_EXT}
First, we consider the PASEP with {\it pw} disorder as defined above Eq.(\ref{delta_p}) with a
special form of the bimodal disorder, when the particles are of two kinds. For a fraction of $c \ll 1$
{\it black} particles the hopping rates are: $p_i=1$ and $q_i=\lambda$, whereas the
$(1-c)$ fraction of {\it white} particles have hopping rates: $p_i=1$ and $q_i=\lambda^{-1}$.
In order to obtain exact results we take the limit $\lambda \gg 1$. With this type of disorder
the drift of the white particles to the right is slowed down by the black ones and the slowing dawn
is even more effective if two or more black particles happen to stay behind each other. A cluster
of $n$ black particles has a very small density: $\rho(n)=c^n$ and its speed to the right, in the
absence of another black particles and clusters is given by, $v(n) \approx \lambda^{-n}$. Indeed in order to make the complete black cluster one step to the
right all the $n$ particles should perform their hop to the unpreferential direction at the same timestep.
At this point it is easy to notice the isomorphism of this problem with the RTIM in Sec.\ref{sec:RTIM_extr}
with the correspondence: $\epsilon \leftrightarrow v$. Thus we can immediately write for the
distribution of the effective velocities of the black clusters:
\be
P(v)=\frac{1}{\ln \lambda} v^{\omega},\quad v \to 0\;,
\label{v_extr}
\ee
with the exponent, $\omega$, defined in Eq.(\ref{omega_extr}).

Evidently, the stationary velocity of the PASEP is given by the smallest effective speed of the clusters, $v_1$,
and all the particles move behind that slowest black cluster. In a large finite system with a finite density
of particles,
the stationary velocity goes to zero as given in Eq.(\ref{v}) with a dynamical
exponent defined in Eq.(\ref{z_extr}). Consequently in a finite system the distribution of $v_1$ is given
in the scaling form in Eq.(\ref{P_L}), and the distribution function of the scaling variable: $u=u_0 v_1 L^{z_p}$ is given by the Fr\'echet distribution in Eq.(\ref{frechet}).

The results obtained in this section can be easily generalized for the PASEP with site-wise ({\it sw}) disorder,
in which case the hop rates depend on the position: for the site $i$ they are $p_i$ (right) and $q_i$ (left).
The control-parameter of the model is the same as for {\it pw} disorder in Eq.(\ref{delta_p}). For the extreme
binary disorder considered above most of the sites are {\it white} and promote the movement of the
particles to the right, but the few {\it black} sites and in particular the rare black clusters form
barriers, which slow down the particle motion. This problem is studied in more details in
Ref.[\onlinecite{pasep_sw}]. Here we just note that the average velocity of a particle which goes
trough a single
large barrier of size, $n$, is given by\cite{bece00}, $v \approx \lambda^{-n/2}$, since due to particle-hole symmetry in the
stationary state the barrier is filled up to $n/2$, so that the particles should make $n/2$ consecutive
steps against the barrier. As a consequence the derivation in the previous paragraphs for {\it pw} disorder
should by slightly modified, which leads to a dynamical exponent, $z_s$, given by:
\be
z_s=\frac{z_p}{2}\;.
\label{z_s}
\ee
In particular the distribution of the stationary velocity in a finite system is still given by the Fr\'echet
form in terms of the scaling variable: $u=u_0 v_1 L^{z_s}$.

\subsection{Strong disorder RG and scaling results}
\label{sec:PASEP_RG}
Here we show that the results in the previous section, i.e the relation between Griffiths singularities
and EVS holds for a general form of disorder, too. We start with the PASEP with {\it pw} disorder for which a
variant of the strong disorder RG approach has been applied\cite{ASEP}. For this model during renormalization
the fastest hop rates are consecutively decimated out and new clusters of particles are created with
effective hop rates obtained by a perturbation calculation. Without going to the details we mention
that there is a one-to-one correspondence between the RG rules for the RTIM and that of the PASEP.
In the Griffiths phase for $\delta_p>0$ in the first part of the renormalization both left and right
hop rates are decimated out until particle clusters of typical size, $\xi$, are created. (Here $\xi$ is
the average correlation length, which behaves for small $\delta_p$ as $\xi \sim \delta_p^{-2}$.) At this scale
of the RG the remaining effective particles have a practically vanishing left hop rate, $\tilde{q}_i \approx 0$,
and finite effective right hop rates, which have an asymptotic power-law distribution:
$P(\tilde{p}) \sim \tilde{p}^{-1+1/z_p}$. Here the dynamical exponent, $z_p$, is given through the
equation:
\be
\left[ \left( \frac{q}{p} \right)^{1/z_p}\right]_{\rm av}=1\;,
\label{z_p1}
\ee
which makes the analogy with the RTIM complete, see in Eq.(\ref{z1}). The renormalized PASEP than
consists of independent particles the velocity (right hop rate) of them has the same power-law
distribution and the smallest of them is the stationary velocity of the finite system. Consequently
the conditions of EVS are asymptotically satisfied, so that the distribution of the stationary
velocity in a finite system is given by the Fr\'echet
form in terms of the scaling variable: $u=u_0 v_1 (L/\xi)^{z_p}$.

For the PASEP with {\it sw} disorder one can not simply apply the strong disorder RG, so that we use
here phenomenological, scaling considerations. As noticed in Sec.\ref{sec:PASEP_EXT} the
rare regions are represented by large barriers, which in the length-scale, $\xi$, are
expected to be independent and well separated from each other. Scaling consideration in Ref.[\onlinecite{ASEP}]
show that the distribution of the velocities associated to large barriers is given by, $P(v) \sim v^{-1+1/z_s}$
and for the dynamical exponent, $z_s$, the relation in Eq.(\ref{z_s}) holds in the entire Griffiths phase.
Once more the stationary velocity in a finite system is given by the smallest velocity associated to
the largest barrier, and its distribution is expected to be in the Fr\'echet
form in terms of the scaling variable: $u=u_0 v_1 (L/\xi)^{z_s}$.

Finally, we consider the 1d zero-range process (ZRP) with quenched disorder\cite{ZRP}.
In this model the $i$-th lattice
site can be occupied by $n_i \ge 0$ particles, from which the topmost one can hop to nearest neighbor
sites with a position dependent rate: $q_{i}$ to site $i+1$ and $p_{i-1}$ to site $i-1$. It is known
that the ZRP with this definition can be exactly mapped (up to translations of the configurations
of the lattice) to the PASEP with {\it pw} disorder, as studied here. The sites of the ZRP are particles in
the ASEP and the particle clusters in the ZRP correspond to holes in front of the particles in the PASEP.
Then the stationary current in the ZRP is just the stationary particle velocity of the PASEP. From this
mapping and the previous reasoning follows that the stationary current of the ZRP, $J$, in a finite
system scales as: $J \sim L^{-z_p}$, and its distribution is the Fr\'echet distribution.

\subsection{Numerical results}

\begin{figure}[!h!]
\begin{center}
\includegraphics[width=3.2in,angle=0]{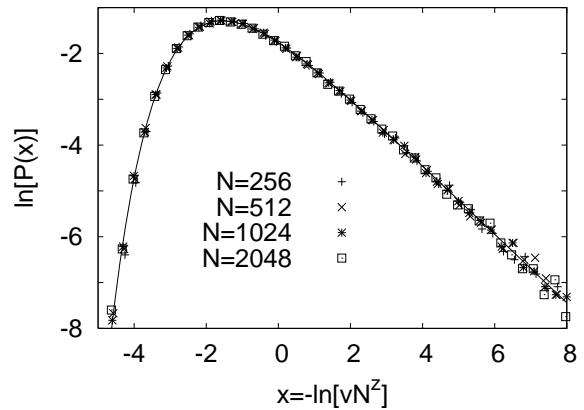}
\end{center}
\caption{Distribution of the stationary velocity of the PASEP with {\it pw} uniform disorder having $h_0=3.$
calculated for different numbers of particles. The Fr\'echet distribution with the exact dynamical exponent in Eq.(\ref{z_p1}) is indicated by the full line.}
\label{fig:PASEP_PW}
\end{figure}

We start to analyze the distribution of the stationary velocity, $v$, of the PASEP with uniform
{\it pw} disorder, which is presented in Fig.\ref{fig:PASEP_PW}. Here we have evaluated an exact
expression for $v$, which is described c.f. in Ref.[\onlinecite{ASEP}] and in this way
we have studied chains with up to $N=2048$  particles over 100000 independent realizations of
the disorder. Fig.\ref{fig:PASEP_PW} shows an excellent agreement with the Fr\'echet distribution
having the exact dynamical exponent in Eq.(\ref{z_p1}).


\begin{figure}[!h!]
\begin{center}
\includegraphics[width=3.2in,angle=0]{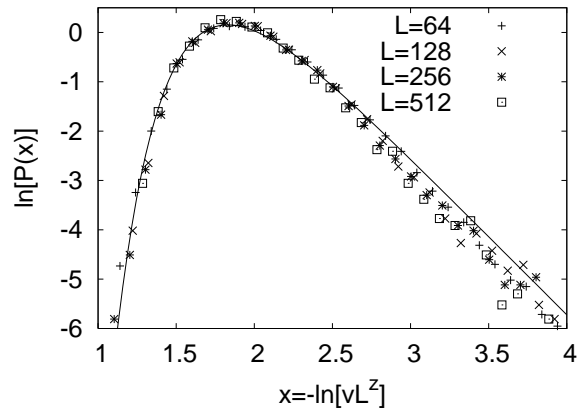}
\end{center}
\caption{The same as in Fig.\ref{fig:PASEP_PW} for the PASEP {\it sw} binary disorder with $c=0.25$
and $\lambda=2.$ For the Fr\'echet distribution, which is indicated by the full line the
dynamical exponent is taken from the
scaling result in Eqs.(\ref{z_s}) and (\ref{z_p1}).}
\label{fig:PASEP_SW}
\end{figure}

For the PASEP with {\it sw} disorder there is no analytical expression for the stationary velocity
so that $v$ is calculated by simulation. We have considered 10000 random half-filled chains with binary
disorder of different lengths up to $L=512$. The results as presented in Fig.\ref{fig:PASEP_SW}
are in good agreement with the Fr\'echet distribution, in which
the exact dynamical exponent is taken from the scaling result in Eqs.(\ref{z_s}) and (\ref{z_p1}).


\section{Discussion}
\label{sec:disc}
In this paper we have considered strong Griffiths effects in different interacting
many particle systems and
studied their possible relation with extreme value statistics. Our examples included random
quantum systems in one and two dimensions as well as stochastic systems with quenched disorder
in one dimension. Our exact, numerical and RG results indicate that for systems
having a discrete symmetry the distribution of the
inverse time scales (excitation energy for quantum systems, stationary velocity for exclusion
models) in large finite samples has a universal form, which is the limit distribution of
the extremes of {\it iid} random numbers. In these examples the Griffiths singularities are characterized
by the dynamical exponent, $z$, which is a continuous function of the control parameter, $\delta$.
However the distribution function depends on, $u=L^{z/d}\tau^{-1}$, and given by
the Fr\'echet distribution in Eq.(\ref{frechet}). The physical picture behind this
result is given by the strong disorder RG method: during renormalization fast degrees of freedom
are gradually decimated out and the system finally transforms
into a set of practically independent degrees of freedom. The characteristic
time-scales of these localized units follow a power-law distribution with a $z$-dependent decay
exponent. Since the physically relevant relaxation time is given by the largest time-scale we arrive
to the results of EVS.

The universality of the distribution function of different problems is
tested by numerical and RG calculations. For models with a discrete symmetry in all cases we
obtained convincing evidence of universality. On the contrary for the random Heisenberg model, which has
a continuous symmetry, the distribution function is found to depend on the specific form of the
disorder. Interestingly, for the 2d spin-glass model as well as for the diluted 2d model above the percolation
threshold the distribution function is found in universal Fr\'echet form.

One might ask the question how general these results are. On the basis of the RG approach we conjecture
that for all interacting systems which have a disordered Griffiths phase the singular properties of which are controlled by the same type of {\it strong disorder fixed points} as for the RTIM the distribution
function of the inverse time-scales is universal. Possible systems of this class are, besides
random quantum magnets and exclusion processes, some reaction-diffusion models\cite{contact}, the dynamics of
the random-field Ising chain\cite{rfim}, the localization of a random
polymer at an interface\cite{polymer}, etc.

Evidently the above considerations of EVS does not apply for the distribution of {\it average} and 
{\it local} physical quantities in random systems. For the RTIM average quantities are, among others the
uniform susceptibility or the sound velocity\cite{luck}, whereas examples for local quantities are
the surface susceptibility or the surface magnetization in the ordered
Griffiths phase\cite{bigpaper,dharyoung,cecile}. The biased random walk in a random
environment\cite{walk} is a one particle problem, thus the applicability of the
EVS is not expected. Indeed, the stationary velocity is an average quantity, since the time needed for the
particle to get through a system is obtained
by averaging the waiting times associated at different points of the lattice\cite{derrida}. 
Therefore the distribution of the stationary velocity is not in the Fr\'echet form\cite{walk}.

\begin{acknowledgments}

We acknowledge useful discussions with C. Monthus, H. Rieger and L. Santen.
This work has been
supported by a German-Hungarian exchange program (DAAD-M\"OB), by the
Hungarian National Research Fund under grant No OTKA TO34138, TO37323,
TO48721, MO45596 and M36803. RJ acknowledges support by the Deutsche Forschungsgemeinschaft
under Grant No. SA864/2-1. 
\\
\end{acknowledgments}


\begin{thebibliography}{}
\bibitem{yr96} This scaling form in the general case follows from the assumption that the
rare regions are localized, thus for a small $\epsilon_1$ we have $P_L(\epsilon_1) \sim L \epsilon_1^{1/z}$.
A.P. Young and H. Rieger, Phys. Rev. B {\bf 53}, 8486 (1996).
\bibitem{XXX} At the particular value, $q=4$, the RG equations are equivalent to that of the
dimerized random antiferromagnetic Heisenberg chain, given in Eq.(\ref{eq:SG}) with $p=0$ and
with different antiferromagnetic couplings at odd and even bonds, see in Ref.\cite{ijl}.
\end{thebibliography}
\end{document}